%% file: acl_main.tex
\documentclass[11pt]{article}
 
\usepackage[preprint]{acl}

\usepackage{times}
\usepackage{tabularx}
\usepackage{latexsym}
\usepackage{booktabs}
\usepackage{multirow}
\usepackage{enumitem} 
\usepackage{makecell}
\usepackage{tcolorbox}
\usepackage{fancyvrb}
\usepackage{amsmath}
\usepackage{listings}
\usepackage[T1]{fontenc}

\usepackage[utf8]{inputenc}

\usepackage{microtype}

\usepackage{inconsolata}

\usepackage{graphicx}

\title{AI Exposure and the Changing Responsibilities in Job Postings}

\newcommand{\aspace}{\hspace{1em}}

\author{
    Dora Zhao\textsuperscript{3}\thanks{Work done while interning at Microsoft.}, 
    Lin Ai\textsuperscript{1},
    Pranav Khadpe\textsuperscript{1},
    Siddharth Suri\textsuperscript{2},
    Scott Counts\textsuperscript{1} \\
    \textsuperscript{1}{}Microsoft\aspace
    \textsuperscript{2}{}Microsoft Research\aspace
    \textsuperscript{3}{}Stanford University  
}

\begin{document}

\raggedbottom
\maketitle

\begin{abstract}
The growing adoption of artificial intelligence (AI) in the workplace has raised questions about how employer demand for different work activities is changing.  We examine how the prevalence and content of work activities in online job postings change in association with activity-level measures of AI exposure, derived from Copilot and Claude usage. To enable systematic comparisons, we develop a pipeline that extracts the responsibilities from job advertisements and maps them onto O*NET's standardized taxonomy of Intermediate Work Activities (IWAs). We then apply this approach to 50,000 job advertisements from 33 firms between 2023 and 2025. From this analysis, we first find that explicit mentions of AI increased by 66\%, although much of this growth occurs outside descriptions of the actual work activities associated with a job. Looking specifically at the advertised job responsibilities, activities with greater AI exposure tend to decline in advertised prevalence, but only mildly as a share of the total number of responsibilities. At the same time, more AI-exposed work activities increasingly are associated with explicitly AI-related responsibilities, suggesting that employers are looking to hire people who can perform these responsibilities with AI. 
\end{abstract}

\section{Introduction}
\input{Sections/01_Introduction}

\section{Related Work}
\input{Sections/02_RelatedWork}
\section{Background}
We start by defining two key terms used throughout our analysis.
\paragraph{Work Activity.} O*NET organizes work activities hierarchically at three levels of granularity: (1)~Generalized Work Activities (GWAs; ``Guiding, Directing, and Motivating Subordinates''), (2)~Intermediate Work Activities (IWAs; ``Coordinate group, community, or public activities.''), and (3)~Detailed Work Activities (DWAs; ``Coordinate special events or programs''). Below these activities are individual tasks (``Organize or approve promotional campaigns''), which represent the most specific unit in the hierarchy. There is a many-to-many mapping between tasks and DWAs. Each DWA, however, belongs to a single IWA, and each IWA to a single GWA. In this work, we use ``work activity'' to refer to IWAs, which provide an intermediate level of abstraction that captures meaningful categories of work while remaining comparable across occupations. We selected this operationalization as prior work analyzing economic impacts of LLMs has also used IWAs~\cite{tomlinson2025working,anthropic2026aeiv6,counts2026ai}, providing a shared vocabulary.

\paragraph{Occupation Group.} We use the Standard Occupational Classification (SOC) system from the U.S. Bureau of Labor Statistics to define occupation groups. SOC organizes occupations hierarchically, covering 23 major groups (``Sales and Related Occupations''), 98 minor groups (``Retail Sales Workers''), and 459 broad occupations (``Cashiers''). We use the term ``occupation group'' to refer to the SOC major groups, selected to ensure sufficient coverage of job postings across the different categories.
\section{Method}
\label{sec:method}
\input{Sections/03_Method}
\input{Sections/04_Findings}
\section{Discussion and Conclusion}
\input{Sections/05_Discussion}

\newpage
\section*{Limitations}
We discuss several limitations with this work. First, it is important to note that our analyses identify \emph{associations}, not causal effects of AI use on advertised work activities. There are other changes during the study period, such as shifts in hiring priorities, firm strategies, or broader macroeconomic events that could contribute to the patterns that we observe. While we have identified that this negative association exists and shown its robustness through using two different operationalizations of AI exposure, we still need further work to uncover the underlying mechanism explaining why this change occurs. 

Second, we discuss limitations related to our measures. Our dependent variable, advertised activity frequency, captures the share of a firm's postings that mention a given activity rather than the amount of work devoted to that activity. This share can decline when the mix of roles being advertised changes, even if demand for workers who perform the activity remains unchanged. Likewise, a mention does not indicate how central an activity will be to an employee's role or how much time they will spend performing it. Our activity frequencies depend on mapping responsibilities described in job postings to O*NET work activities. Because job postings describe similar responsibilities in different ways and our mapping is imperfect, some of the variation we observe may reflect differences in how work is described rather than differences in the work itself.

In addition, our measures of AI exposure also capture only a subset of how AI is used in the workplace. Specifically, they measure the extent to which people use general-purpose LLM chatbots, such as Copilot or Claude, to perform different work activities. However, they do not capture exposure to other forms of AI, including agentic systems (e.g., coding agents) that have become popular for tasks such as software engineering, more embodied forms of AI (e.g., robots, autonomous driving), or specialized AI embedded within industry-specific tools. As these technologies diffuse unevenly across occupations and activities, the relationships we observe should be interpreted as changes associated with exposure to general-purpose LLM use rather than to AI or automation more broadly.

Finally, our sample is drawn from larger U.S.-based firms and is concentrated in knowledge-work occupations. The patterns we observe may differ at smaller firms, outside the United States, or in occupations less well represented in our data. Our exposure measures also capture activity-level use of Copilot and Claude. While these cover popular LLMs used in the U.S., it is possible that users of these products may differ from workers at the sampled firms. 

\section*{Ethical Considerations}
A key consideration around this work is related to how our findings are interpreted. This work focuses on how work activities are advertised in job postings, rather than changes in employment or the work employees actually perform. Given the significant public interest in AI's effect on the workforce, an important ethical consideration is overinterpretation of our results to make any claims about how AI affects hiring patterns. As stated throughout the paper, job postings reflect the advertised requirements of employers, not hiring outcomes. Furthermore, our work only provides correlations between patterns in job postings and AI exposure, not causal evidence. Overstating these findings could contribute to misleading narratives about worker displacement or the future demand for particular occupations. 
\bibliography{egbib}
\appendix
\label{sec:appendix}
\input{Sections/999_Appendix}

\end{document}

%% file: Sections/01_Introduction.tex
Generative AI has rapidly diffused into the workplace since the release of ChatGPT in November 2022. Nationally representative U.S. surveys conducted in August and November 2024 found that 23\% of employed adults aged 18–64 had used generative AI for work in the previous week, including 9\% who used it every workday \cite{bick2026rapid}. Analyses of real-world AI usage further document its extensive application to knowledge-work activities \cite{tomlinson2025working, counts2026ai, anthropic2026aeiv6}. Yet evidence about workers’ use of AI does not by itself reveal how the work employers demand is changing. For example, as AI becomes integrated into workplaces, employers may seek greater AI-related capabilities while also changing the mix of other activities they expect workers to perform. Thus, we extend this literature by examining changes in the composition of advertised work associated with AI exposure. 

\begin{figure*}
    \centering
    \includegraphics[width=\linewidth]{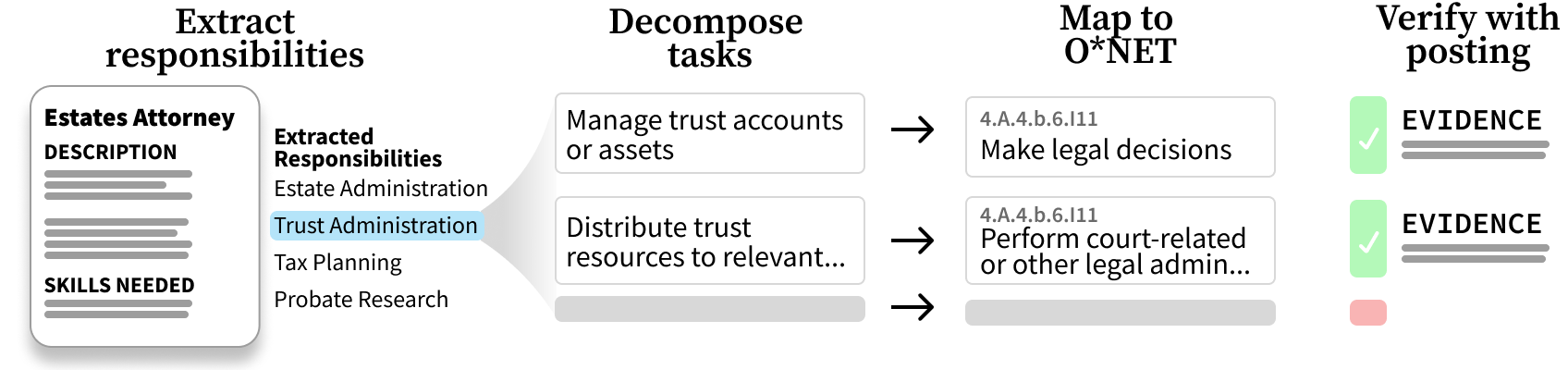}
    \caption{We present a method for creating a standardized representation of work activities described in a job posting. First, we extract the job title and listed tasks from the postings using an LLM. Then, for each job task, we decompose it into granular subtasks, matching the taxonomic structure from O*NET. For each of the decomposed tasks, we then select the most similar O*NET work activity based on embedding cosine similarity and verify mappings using a separate LLM. We also map the extracted job title in each posting to an occupation in O*NET.}
    \label{fig:pipeline}
\end{figure*}

Job postings offer one way to study changes in employers' demand for different types of work~\cite{squicciarini2021demand,green2024artificial}. Recent analyses found that the share of U.S. job postings mentioning AI has tripled since 2022~\cite{stahle2026jobsai}, with this increase in AI mentions occurring across a range of occupations. But the growing prevalence of explicit AI mentions tells us little about how the broader set of work activities described in job postings is changing, particularly activities that AI can increasingly assist with.

We therefore examine two related questions. First, we look at how the prevalence of advertised work activities is changing in relation to their exposure to AI, and second, how AI is incorporated into the responsibilities associated with those activities. For instance, an activity may become less prevalent in job postings overall while an increasing share of postings containing that activity describes it in AI-related terms. At the same time, changes in advertised prevalence may reflect not only changing demand for work but also changes in how employers describe and consolidate responsibilities. Our goal is thus to characterize how the composition and content of advertised work are evolving alongside the diffusion of AI.

To analyze these changes, we require a method that converts job postings, which describe work using heterogeneous language and at varying levels of specificity, into a standardized representation. We develop a pipeline that extracts tasks from postings and maps them onto O*NET's Intermediate Work Activities (IWAs), a taxonomy maintained by the U.S. Department of Labor (see Fig.~\ref{fig:pipeline}). This representation enables comparisons of postings across time periods and firms.

We apply this method to analyze 50,000 U.S. job postings from 33 firms, posted between 2023 and 2025. Our analysis yields three main findings. First, we look at the \textbf{visibility} of AI in postings, corroborating that explicit mentions of AI are increasing (Sec.~\ref{sec:visible}). However, these mentions do not always describe new work responsibilities; AI also increasingly appears in how firms describe their products, organizational capabilities, and expected qualifications, rather than solely in the tasks workers are expected to perform. Second, activities with greater AI exposure tend to decline in their advertised \textbf{prevalence} (Sec.~\ref{sec:prev}). The strength of this association varies across measures of prevalence, with some of the overall decline reflecting changes in how firms describe and enumerate responsibilities in their postings. Finally, in Sec.~\ref{sec:content}, by looking at the substantive \textbf{content} of these postings, we find that more AI-exposed activities are increasingly associated with AI-related responsibilities, even as their overall advertised prevalence declines ($\beta=0.28$, $p=.001$). Looking at activity prevalence alone would overlook the growing AI-related responsibilities within existing activities.



%% file: Sections/02_RelatedWork.tex
\paragraph{Measuring Worker Use of AI.} One thrust of prior work measures how people are using AI in the workforce. Chatbot providers have released industry reports characterizing the economic tasks for which their products are used, providing empirical insight into what activities workers deploy LLMs for~\cite{handa2025economic,tomlinson2025working,counts2026ai,chatterji2026organizations} and misalignments between existing capabilities and workers' needs~\cite{shao2025future,wang2026well}. These efforts have also been paired with concurrent work from AI developers to evaluate these technologies on economically valuable tasks that better reflect how they are used in the workplace~\cite{patwardhan2026gdpval,haupt2025position}. However, patterns of worker AI use capture only one side of how AI may be reshaping work, as what workers choose to do with these technologies does not necessarily reflect how employer demand is changing. By examining job postings, our work focuses on the other side of this equation. Measuring what work activities employers advertise for reveals how they expect job tasks to change as AI grows in popularity, independent of whether workers have adopted these technologies in practice. 

\paragraph{Evaluating the Economic Impact of LLMs.}
Rather than measuring how workers are using AI, another body of work has focused on understanding how AI is reshaping the work being done. Taking a broader view, prior work has aimed to assess which jobs are most likely to be affected by AI, characterizing the degree of exposure across occupations~\cite{webb2019impact,eloundou2024gpts,felten2019effect}. Others have sought to quantify the effects of integrating AI into the workplace on factors such as productivity~\cite{becker2025measuring,noy2023experimental,brynjolfsson2025generative,dell2023navigating} and skill development~\cite{shen2026ai,brynjolfsson2025generative}. A natural next question is \emph{how} the requirements of these jobs are changing. Recent work has begun to address this question by examining changes in the skills and tasks employers advertise, finding shifts in skill demand associated with AI exposure and adoption~\cite{green2024artificial,gulati2024mitigating,gulati2025generative}. Other work has compared labor demand for work that is more amenable to substitution versus augmentation by generative AI, finding heterogeneous effects across different types of work~\cite{chen2025displacement,teutloff2025winners}. More recently, \citet{wang2026generative} examined whether changes in AI exposure arise through shifts in the jobs firms hire for or changes in the tasks contained within jobs. This literature provides early evidence that the diffusion of AI is associated with changes in the composition of labor demand. We build on this work by focusing on how the specific work activities employers advertise change over time in relation to observed patterns of AI use.

\paragraph{Measuring Labor-Market Demand from Job Postings.}
Since job postings provide a valuable lens into labor-market change, a growing body of NLP research has developed methods for extracting skill requirements from job advertisements~\cite{zhang2022skillspan,gnehm2022fine,senger2024deep,nguyen2024rethinking}. We build on this work by leveraging the capabilities of LLMs to map the responsibilities described in job postings onto standardized O*NET work activities. Furthermore, we use the representation produced by our pipeline as a measurement instrument to conduct a series of empirical analyses focused on how employers' expected requirements have shifted over time.

%% file: Sections/03_Method.tex
\begin{table*}
\footnotesize
    \centering
    \begin{tabular}{@{}p{0.25\textwidth} p{0.6\textwidth} p{0.075\textwidth}@{}}
    \toprule
    Industry & Companies (\# of Employees) & Postings \\
    \midrule
    Information & Microsoft ($\sim$228K), Comcast ($\sim$179K), Salesforce ($\sim$76.5K), Nexstar Media Group ($\sim$13K) & 5,773 \\
    Manufacturing & General Motors ($\sim$156K), 3M ($\sim$60.5K), Watts Water Technologies ($\sim$5K) & 5,600 \\
    Retail Trade & Five Below ($\sim$18K), Dollar General ($\sim$194K), Walmart ($\sim$2.1M) & 5,600 \\
    Finance \& Insurance & Bank of America ($\sim$213K), JPMorganChase ($\sim$318.5K), PNC Financial Services Group ($\sim$55.3K), Morgan Stanley ($\sim$83K) & 5,508 \\
    Mining, Quarrying, Oil \& Gas & Freeport-McMoRan ($\sim$28.5K), ExxonMobil ($\sim$58K), Cleveland-Cliffs ($\sim$25K) & 5,179 \\
    Utilities & Evergy ($\sim$4.7K), Xcel Energy ($\sim$11.4K), Duke Energy ($\sim$26.4K)& 5,085 \\
    Transportation \& Warehousing & UPS ($\sim$460K), XPO ($\sim$38K), Old Dominion Freight Line ($\sim$24K) & 4,927 \\
    Wholesale Trade & McKesson ($\sim$45K), Wesco ($\sim$20K), ScanSource ($\sim$1.4K) & 4,859 \\
    Construction & Primoris Services Corporation ($\sim$12K), Fluor Corporation ($\sim$23K), Tutor Perini Corporation ($\sim$7.5K) & 4,262 \\
    Agriculture, Forestry, Fishing \& Hunting & Corteva Agriscience ($\sim$21.5K), Fresh Del Monte ($\sim$34K), Cal-Maine Foods ($\sim$4K), Chiquita Brands ($\sim$20K) & 3,207 \\
    \bottomrule
    \end{tabular}
    \caption{We sample job postings from 33 companies stratified by sector. Our sampling targeted at least three companies per sector, yielding 3,207--5,773 postings per sector. For each sector, we report the selected companies and the approximate number of employees at the company at the time of sampling.}
    \label{tab:companies}
\end{table*}

Job postings include textual descriptions of what the key tasks are for a given role. However, postings are not standardized, differing in terms of format, description granularity, and so on. In this section, we introduce our method for creating a standardized representation of tasks in a job posting and its accompanying validation.

\subsection{Vectorizing Job Postings}
Our pipeline takes in as input a job posting and outputs a vector with 333 dimensions. Of these, 332 correspond to work activities in O*NET and one corresponds to the matched occupational title.\footnote{We use O*NET 30.2 released in February 2026~\cite{onet_database}, and the 2018 Standard Occupational Classification System.}

As shown in Fig.~\ref{fig:pipeline}, the first step in our pipeline is to \textbf{extract the job responsibilities}. Since job postings are not uniform in how they are structured, we use an LLM (\texttt{GPT-5.1}) to extract the set of discrete responsibilities and tasks described in the posting, thereby producing a list of advertised responsibilities and duties associated with the role. 

Our second step is to \textbf{standardize tasks} by decomposing each one into a set of smaller, more atomic tasks, again using \texttt{GPT-5.1}. Advertised job responsibilities are not consistent in granularity. For example, the job posting for an estate associate attorney contains broad tasks, such as ``trust administration'' and ``estate administration'' --- each of which encompasses many more granular work activities. Our decomposition is guided by the taxonomic structure used in O*NET, ensuring that the resulting sub-tasks are expressed at a level of specificity that facilitates downstream matching.

Once we have the extracted information from the job description, we then match the tasks and job title to O*NET. Following prior work~\cite{tomlinson2025working}, we \textbf{match each decomposed task} to entries in the O*NET taxonomy using embedding-based similarity (\texttt{text-embedding-3-large}), selecting the task that has the highest cosine similarity. Each posting is represented as a distribution over IWAs by counting the IWAs across all of its tasks and normalizing to sum to one. This pipeline can be easily adapted to map to any activity in the O*NET hierarchy (e.g., DWA, GWA). We run the decomposition and matching process twice, taking the union of the proposed IWAs. Finally, we \textbf{verify} that there is evidence for the IWA in the job description using \texttt{GPT-6 Luna}. The verifier returns a binary judgment as to whether the IWA should be included and evidence that cites the specific text in the job requirement supporting this decision.

We also \textbf{map the job title} listed in each posting to an O*NET occupation again using embedding-based similarity. The title provides additional metadata about the role and enables flexible mapping from O*NET occupations to the Standard Occupational Classification (SOC) system. 

\paragraph{Pipeline Validation.}
We first validate that the IWA extraction pipeline performs accurately end-to-end, assessing whether it can assign correct work activities given a job posting. In this validation task, three authors independently reviewed 10 postings, each paired with a candidate set of IWAs. For each posting, half of the candidate IWAs were identified by our pipeline, while the other half were distractors that were not present. Authors selected all candidate IWAs represented in the job description. Each author evaluated 180 IWA–posting pairs, yielding 540 binary judgments in total. Inter-annotator agreement was moderate (average Cohen's $\kappa$ = 0.53), consistent with prior work which has pointed out the inherent difficulty of this task~\cite{tomlinson2025working}. Using majority vote across annotators as ground truth, our pipeline achieves a precision of 0.91 and recall of 0.62. Additional ablation analyses reported in Appendix~\ref{sec:app_pipeline}. 

Next, we evaluate the accuracy of our title-matching procedure. Authors were shown the original job title alongside the matched O*NET title and asked whether the match was exact. When the title was not an exact match, they were asked to identify whether a better-matching O*NET title existed. Doing so allowed us to separate errors attributable to our classifier from limitations inherent to the O*NET taxonomy itself. Annotators labeled 50 job titles in total (mean Cohen's $\kappa = 0.86$). We report results on the 40 titles for which annotators reached consensus on the best-matching title. On this set, the pipeline achieves exact-match accuracy of $72.5\%$ and correctly identifies the broad occupational group $77.5\%$ of the time.

\subsection{Data}
\paragraph{Sampling Strategy.} To ensure variation across the companies we study, we purposively sampled U.S. companies spanning a range of industries, as defined by North American Industry Classification System (NAICS) codes and company sizes within each industry. Descriptive statistics of the 33 companies are in Table~\ref{tab:companies}. We derive each company’s NAICS code from its self-reported industry classification in the SEC EDGAR database. EDGAR reports a Standard Industrial Classification (SIC) code, which we map to the corresponding NAICS code using the official SIC–NAICS crosswalk~\cite{schaller2021weighted}. For each company, we construct a ``near-universe'' of job postings from 2023Q1-2025Q4 using job postings data from the platform TheirStack~\footnote{https://theirstack.com/en/job-posting-api}, which aggregates and deduplicates job listings across the web. The platform's public API allows us to retrieve all available job postings for a given company, including historical postings. 

Using this API, we compile 575,686 job postings across the 33 companies. We restrict the dataset to postings in the U.S. with a parseable posting date, leaving 341,220 postings. For our analysis, we sample 50,000 of these postings, stratified by firm and half-year. Since junior- and C-level postings are relatively rare, we retain all postings at these levels. We fill the remainder of the sample by randomly sampling the same fraction of the remaining postings for each firm in each half-year. This procedure preserves each firm’s relative posting volume over time while ensuring sufficient representation across seniority levels.

\paragraph{Metadata.}

We note two important features of this data sample. First, the companies represented are large in size. The smallest company in our dataset, ScanSource, still employs $\sim$1.4K employees. We focus on larger companies given that they comprise an outsized share of labor demand, and prior work provides evidence that adoption of AI in public firms is concentrated in larger companies~\cite{chatterji2026organizations}. Second, even though our dataset is diverse across industries, the occupations in the job postings are concentrated in knowledge-work domains with $16.4\%$ coming from Computer \& Math and $16.1\%$ from Management occupations. Our subsequent analyses are best interpreted as characterizing large firms and knowledge-work occupations specifically.


%% file: Sections/04_Findings.tex
\label{sec:valid}

\section{Growing Visibility of AI in Job Postings}
\label{sec:visible}
We begin by analyzing the visibility of AI in job postings, measuring how frequently employers explicitly mention AI.

\subsection{Analysis Method} This analysis uses the raw text of each job description rather than the extracted IWAs. We flag a posting as mentioning AI if its description contains an AI-related term (e.g., ``machine learning,'' ``LLM,'' or ``generative AI''). Since many firms add the same AI-related language to postings across roles (e.g., "Salesforce is the \#1 AI CRM"), we exclude boilerplate sentences that appear in at least 10\% of a firm's postings within a half-year across at least five distinct job titles. To estimate the annual change in the probability that a posting mentions AI, we use a linear probability model with firm and occupation-group fixed effects, clustering standard errors by firm. Additional details are in Appendix~\ref{app:ai-mentions}.


\subsection{Results} 
\paragraph{Mentions of AI are increasing in postings.} In line with existing reports~\cite{stahle2026jobsai,maslej2025artificial}, the overall visibility of AI is increasing in job postings. From 2023 to 2025, the share of postings mentioning AI increased by $66.0\%$, from 5.3\% to 8.8\%. Within firms and occupation groups, the share rose by 2.37 percentage points per year. Specifically, mentions of generative AI increased by $278\%$, from 0.9\% to 3.4\% of postings (see Appendix Table~\ref{tab:ai-mentions-trends} for additional details).


\paragraph{Most growth in AI mentions occurs outside descriptions of work.} The increase in AI mentions does not stem only from changes in the work being described. We label each sentence mentioning AI by the section in which it appears in the job description: responsibilities, qualifications, the summary or opening text, or company context (e.g., company description, values, or benefits). While the number of these sentences mentioning AI more than doubled (2{,}043 to 4{,}792), only 31\% of these sentences were describing the job responsibilities. The rest came from qualifications (29\%), summaries (28\%), and company context (12\%). These sentences often describe skills the applicant should bring (e.g., ``Awareness of generative AI tools (e.g., ChatGPT, Microsoft Copilot) and how to use them to enhance productivity'') or present the company as AI-oriented (e.g., ``Join Salesforce and define the future of Customer Relationship Management Solutions with a strong emphasis on AI-driven innovation''). Mentions of AI in job postings therefore reflect not only how work is changing but also the skills employers seek and how firms present themselves to applicants.



\section{AI Exposure and Changes in Advertised Activity Prevalence}
\label{sec:prev}
Next, we look at the changing prevalence of AI-exposed work activities. Building on prior work that examines the relationship between exposure and employer demand~\cite{munyikwa2025tasks,green2024artificial}, we extend this analysis to a more recent time period and across multiple measures of activity prevalence. Doing so allows us to distinguish declines in the presence of individual activities from shifts in the relative composition of advertised work, while also assessing whether changes in how employers describe job responsibilities contribute to observed declines.

\subsection{Analysis Method} We use our dataset of job postings, selecting those that fall into two windows. The first, $T_1$, includes job postings from 2023 and the second, $T_2$, from 2025.\footnote{We use 2023 and 2025 as they are the earliest and latest years in our dataset for which we have job postings from the entire year. The $T_1$ window includes job postings made soon after the release of ChatGPT in November 2022.} Our first measure is \emph{AI Exposure}, which captures how much existing AI systems are used to complete a work activity. We provide two operationalizations of this variable. From \citet{tomlinson2025working}, we use their measure of the share of a work activity that Microsoft Copilot performs (\emph{Copilot Usage}). From \citet{anthropic2026aeiv6}, we derive \emph{Claude Usage}, the percentage of all Claude conversations mapped to that IWA. We apply a log-transformation to both measures and normalize them. To ensure data coverage, we keep only firms with at least 200 postings in each window, leaving 20 firms and 22{,}963 job postings.\footnote{We use this threshold to reduce noise in the outcome as we are interested in the change in frequency across the two time periods. We report results using alternative thresholds in Appendix~\ref{app:robustness}, finding directionally consistent results.} We exclude IWAs with AI Exposure values below $10^{-6}$, leaving 305 IWAs for Copilot Usage and 154 for Claude Usage.

Our second measure is the change in IWA prevalence between $T_1$ and $T_2$, which we operationalize in three ways:\footnote{We conduct additional robustness analyses reported in Appendix~\ref{app:robustness}.}
\begin{itemize}[leftmargin=*,noitemsep, topsep=1pt]
    \item \textsc{Change in presence.} The share of the firm's postings that mention the IWA in $T_2$ minus the share in $T_1$, in percentage points. 
    \item \textsc{Log ratio.} The log of the IWA's presence in $T_2$ divided by its presence in $T_1$. The log ratio measures relative change in prevalence.
    \item \textsc{Mention share.} The IWA's share of all IWA mentions within the firm in $T_2$ minus its share in $T_1$. This outcome removes overall changes in how many activities postings list and captures only shifts in the \emph{mix} of activities.
\end{itemize}
We fit a linear mixed-effects model pooling across firms, with a random intercept and a random slope on exposure by firm. To examine heterogeneity across firms, we also fit the regression separately for each firm and apply a Benjamini--Hochberg correction across firms.

\subsection{Results}
\begin{figure}
    \centering
    \includegraphics[width=\linewidth]{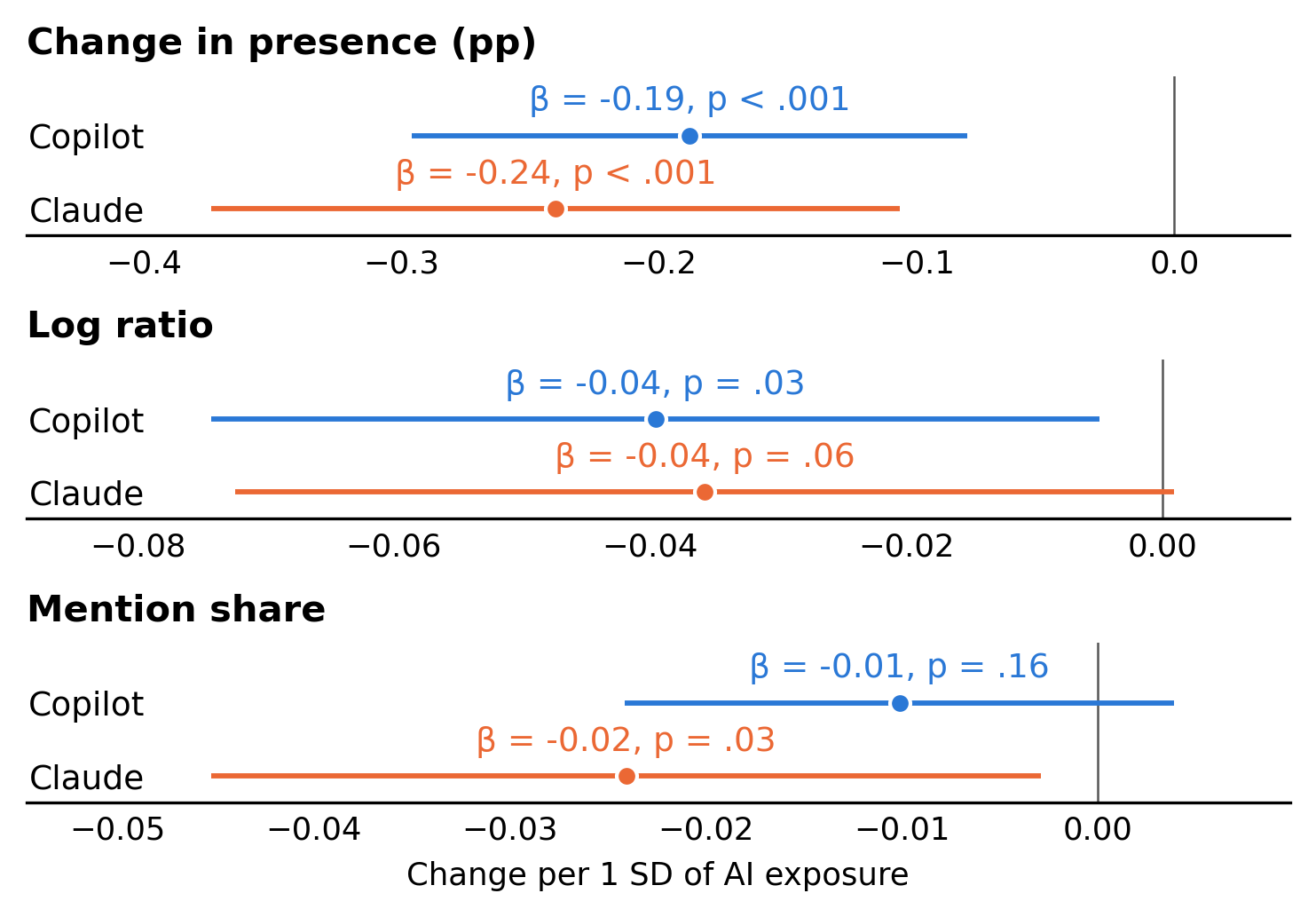}
    \caption{AI exposure is associated with a decline in prevalence from 2023 to 2025. Each point is the slope from a linear mixed model of an activity's change in prevalence within a firm on standardized AI exposure. We provide three definitions of change in prevalence (Change in Presence, Log Ratio, and Mention Share) and two of AI exposure (Copilot Usage, Claude Usage). Horizontal bars are $95\%$ confidence intervals.}
    \label{fig:forest}
\end{figure}
\paragraph{Declines in AI-exposed activities are consistent but sensitive to outcome measurement.} As shown in Fig.~\ref{fig:forest}, across all outcome and exposure measures, the estimated association between AI exposure and change in prevalence is negative, although its magnitude and statistical significance vary across operationalizations. The relationship is strongest for \textsc{change in presence}, where a one-standard-deviation increase in exposure is associated with a 0.19 pp ($p < .001$) larger decline for Copilot Usage and 0.24 pp ($p < .001$) for Claude Usage. On \textsc{log ratio}, the association is smaller but significant for Copilot Usage ($\beta = -0.040$, $p = .03$) and marginal for Claude Usage ($\beta = -0.036$, $p = .06$). Evidence that these activities are declining as a share of all advertised responsibilities is comparatively weaker. On \textsc{Mention Share}, the association is significant for Claude Usage ($\beta = -0.024$, $p = .03$) but not for Copilot Usage ($\beta = -0.010$, $p = .16$). 

\paragraph{Changes in how job descriptions are written complicate prevalence estimates.}
Across time periods, the mean number of extracted responsibilities per posting fell from 12.0 to 10.5, complicating the interpretation of declining activity prevalence. An audit of matched postings (i.e., postings with the same job title and firm across time periods) suggests that much of this decline reflects changes in how postings are written, including revisions to templated responsibility lists (Appendix~\ref{app:taskcount}). This broader decline in listed responsibilities may contribute to the stronger negative associations observed for \textsc{Change in Presence}, which captures both changes in the number and composition of advertised activities. In contrast, \textsc{Mention Share} normalizes by the total number of activity mentions, isolating changes in their relative composition. The weaker associations for this measure suggest that AI-exposed activities are becoming less prevalent in postings, but there is less consistent evidence that they are losing prominence relative to other advertised activities.

\section{Incorporation of AI into Existing Work Activities}
\label{sec:content}
Finally, we turn to examining the content of these work activities, assessing whether employers are increasingly describing existing work activities in connection with AI-related responsibilities.

\subsection{Analysis Method} We trace each sentence that mentions AI in the job posting through our pipeline. A verified IWA is considered ``AI-linked'' if the supporting evidence identified by the verifier in our pipeline either explicitly mentions AI or originates primarily from an AI-related task (see Appendix~\ref{app:ai_linkage}). For each IWA in a given year, we define its \emph{AI-Linked Share} as the percentage of postings containing the IWA in which it is AI-linked. Intuitively, this measures the probability that an IWA is associated with AI in the text of the job posting given that the IWA is present in the posting. 

We relate this share to the work activity's AI Exposure (Copilot Usage), both in 2025 and in its change from 2023 to 2025. We use the same 20 firms as in Sec.~\ref{sec:prev}, keeping IWAs with at least 30 postings in 2025 for the 2025 analysis (219 IWAs) and in both years for the change analysis (206 IWAs). We compute 95\% confidence intervals by resampling firms 1{,}000 times, keeping each firm's 2023 and 2025 postings together.

\subsection{Results}
\begin{figure}
    \centering
    \includegraphics[width=\linewidth]{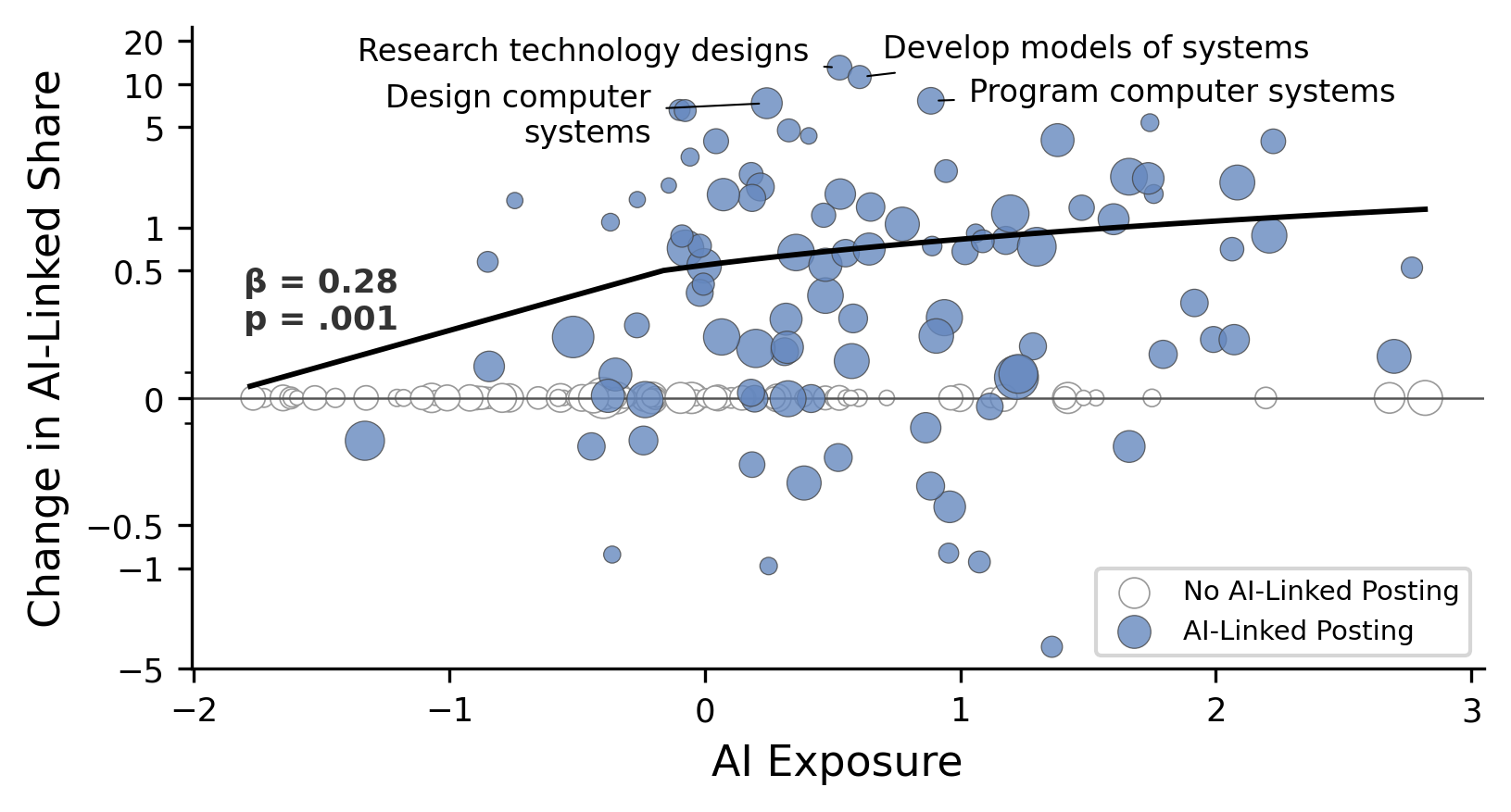}
    \caption{AI-related responsibilities are increasingly concentrated in highly exposed work activities. Each point represents an IWA, plotted by its AI Exposure (Copilot Usage) and the change in AI-Linked Share from 2023 to 2025. Point sizes indicate the number of postings containing each IWA, and black line shows the fitted OLS trend.}
    \label{fig:correlations}
\end{figure}
We start by examining whether AI-related responsibilities are more prevalent among highly exposed work activities and whether this relationship has strengthened over time. Then, we investigate whether the growth in AI-linked responsibilities persists even among activities whose overall advertised prevalence has declined.

\paragraph{More exposed activities are increasingly associated with AI-related responsibilities.} First, we analyze the relationship between AI-Linked Share and AI exposure. In 2025, we find a moderate positive correlation between the AI-Linked Share and how exposed an IWA is (Spearman $\rho = 0.45$, $p =
.001$). AI-Linked Share ranged from just $0.07\%$ among activities in the bottom quartile of AI Exposure to $1.31\%$ for those in the top quartile, suggesting that AI-related responsibilities are disproportionately concentrated in activities already exposed to AI.

Moreover, this concentration has increased over time. As shown in Fig.~\ref{fig:correlations}, between 2023 and 2025, the AI-Linked Share grew more rapidly among exposed activities ($\beta = 0.28$, $p=.001$). Activities in the most exposed quartile experienced a significantly greater increase in AI-Linked Share than those in the least exposed quartile (0.82 pp, $p < .001$). Thus, AI-related responsibilities are not only more common in highly exposed activities but are also growing faster within them.

\paragraph{AI-related responsibilities increase even as activities decline in overall prevalence.} Finally, we find AI-Linked Share increases even among activities that are becoming less prevalent in job postings. For example, the overall presence of ``design computer or information systems'' declined from 9.9\% to 7.2\%, while its AI-Linked Share increased from 2.5\% to 9.8\%. Similarly, ``program computer systems'' declined from 6.1\% to 4.0\% in presence, while its AI-Linked Share rose from 3.1\% to 10.7\%. Additional examples are provided in Appendix~\ref{app:change_analysis}.

An increase in AI-Linked Share could, however, reflect a shrinking number of postings mentioning an activity rather than an increase in postings describing AI-related responsibilities. To distinguish these explanations, we examine changes in the number of postings in which each activity is AI-linked. Among these 57 activities whose overall prevalence declined, nearly all (55 of 57) saw an increase in AI-linked postings; the remaining two showed no change. Even as these activities appear in a smaller share of job postings, they are increasingly advertised in connection with AI-related responsibilities. This pattern highlights a shift in the \emph{content} of advertised activities that measures of prevalence alone would not capture.

%% file: Sections/05_Discussion.tex
We introduce a method for extracting standardized work activities from job postings and apply it to analyze how these advertisements have changed with the popularization of AI in the workplace. We conclude by discussing the implications of this work and identifying directions for future inquiry:

\paragraph{AI is changing the \emph{content} of advertised work activities.}
In this work, we trace how job descriptions are changing as AI becomes more commonplace in the workplace through three lenses: visibility (explicit mentions of AI), prevalence (frequency of AI-exposed work activities), and content (proportion of activities associated with AI-related responsibilities). To the extent that job postings reflect employer demand for different work activities, our findings ostensibly suggest that the demand for activities that can more readily be performed with AI is declining. Yet, when employers do advertise these activities, they are increasingly described in connection with AI-related responsibilities. This pattern suggests that employers are seeking out workers who are either using AI to help complete these activities or are working on AI itself (e.g., developing AI systems). Understanding how employer demand is evolving therefore requires examining both which categories of work appear in job postings and the content of the responsibilities that employers expect workers to do within them.

Our findings also highlight a limitation of relying on standardized occupational taxonomies to understand these fast-paced AI-driven changes in work. Frameworks such as O*NET provide a common vocabulary for comparing work activities, but activities classified under the same category may increasingly entail different responsibilities as AI becomes more integrated in the workplace. Analyses based solely on existing activity categories risk overlooking changes in the content of work within those categories. While there are efforts to update O*NET work activities to account for the diffusion of AI, our work also points to the continued need for research that not only broadens these taxonomies but also provides methods for scalably keeping such taxonomies up to date~\cite{klein2025identification,kell2026modernizing}. Doing so would enable a more complete account of how work evolves alongside AI, including changes that may not be visible through conventional measures of occupational composition.

\paragraph{Job postings reflect both requirements and signaling.}
Interpreting these changes requires considering what job postings represent. Postings communicate the capabilities and responsibilities employers seek for an occupation. At the same time, they are also signals through which firms position jobs and themselves to prospective workers~\cite{connelly2011signaling,gregory2013understanding}. For instance, firms may emphasize AI to signal they are technologically forward-looking, while leaving unstated activities that remain important but are tacit expectations. Changes in advertised activities may therefore reflect changes in the work itself, changes in what employers consider worth stating explicitly, or changes in how firms choose to present jobs. It is also important to note that postings do not reveal who ultimately gets hired or what activities workers actually perform once hired. Future work linking postings to hiring outcomes and workers' realized activities could establish when shifts in advertised requirements translate into changes in the workforce and the work itself.

%% file: Sections/999_Appendix.tex
\section{Pipeline Details}
\label{sec:app_pipeline}
We provide additional details on our method for extracting a standardized work activity representation given a job posting. 

\subsection{Ablation Study}
We conduct the following ablation analyses on our pipeline to validate our implementation: (1)~\textsc{Full}: full implementation of our pipeline; (2)~\textsc{Full - Repeat}: pipeline without the repeated run of the decomposition process; (3)~\textsc{Full - Verification}: pipeline without the verifier; and (4)~\textsc{Full - Repeat - Verification}: pipeline on a single run without verification. As shown in Table~\ref{tab:ablation}, \textsc{Full} offers the best F1 score of $0.74$ with a precision of $0.91$ and recall of $0.62$. Furthermore, compared to the second best-performing implementation --- \textsc{Full - Repeated} which achieves an F1 score of $0.73$ --- \textsc{Full} identifies more IWAs on average with a mean of 8.6 compared to \textsc{Full - Repeated}'s 3.7.

\begin{table*}
    \centering
    \footnotesize
    \begin{tabular}{lrrrr}
    \toprule
         &  F1 & Precision & Recall & Avg. IWAs\\
    \midrule
    \textsc{Full} & \textbf{0.74} & 0.91 & 0.62 & 8.6\\
    \textsc{Full - Repeated} & 0.73 & \textbf{0.92} & 0.61 & 3.7\\
    \textsc{Full - Validation} & 0.63 & 0.52 & \textbf{0.81} & 17.4\\
    \textsc{Full - Repeat - Validation} & 0.63 & 0.52 & 0.79 & 7.5\\
    \bottomrule        
    \end{tabular}
    \caption{The full pipeline outperforms ablated alternatives in terms of F1 score. We evaluate the work activities assigned to a given job posting against our gold set of human annotations.}
    \label{tab:ablation}
\end{table*}

\subsection{Examples}
\input{Figures/JobTitleExamples}
\input{Figures/JobDescriptionExamples}
First, we include examples of outputs from the pipeline described in Sec.~\ref{sec:method}. In Table~\ref{tab:task-mapping-example}, we provide an example of the mapped O*NET tasks based on the original job posting. In Table~\ref{tab:title_examples}, we provide an example of mappings between the occupation titles in the job posting and the selected O*NET occupation. 

\subsection{Prompts}
In our pipeline, we use LLMs for extracting the job description from the posting (see Fig.~\ref{fig:description-extraction-prompt}), standardizing the extracted tasks (see Fig.~\ref{fig:task-standardize-prompt}), and validating tasks (see Fig.~\ref{fig:verifier-prompt}). 

\begin{figure*}[t]
\begin{lstlisting}[basicstyle=\ttfamily\scriptsize, breaklines=true, columns=fullflexible, frame=single, framesep=4pt, xleftmargin=4pt, xrightmargin=4pt]
Your job is to extract all the task descriptions from a job listing as an array. Preserve the description exactly AS-IS in the listing. Do NOT make any changes. If there are no tasks, return an empty array.

Recall that tasks are different from the required abilities for the job. For example, a task is "Travel to client sites to explain documents" whereas an ability is "Ability to travel in the US".

Only include tasks that are specific. Do NOT include catch-all tasks such as "other tasks will be assigned as needed" or "other duties as assigned".
# Examples
## Input
HCC is looking to hire professional and experienced private swim instructors to teach swim lessons to our members at club pool. We supply the clients, you set your own schedule! Must have the ability to work mornings. Conduct private (1-1) and small group swim lessons. Instructor must have their CPR certification. Earn up to $30/hr. Interested parties, email your resume to Estella@harlingencc.com

## Output
{{
    tasks: ["Conduct private (1-1) and small group swim lessons."]
}}

## Input
Growing, boutique law firm seeks a transactional attorney with 3+ years of relevant experience. Based in New York City, we are a close-knit firm with a nationally-recognized franchise and business law practice and an active litigation department. We are currently looking to hire a bright and motivated transactional associate with a strong background representing private companies in areas of corporate law, finance, and mergers & acquisitions. Franchise-specific or commercial real estate leasing experience is not required, but is appreciated. Primary responsibilities will include: Drafting, reviewing and negotiating various corporate and commercial agreements;Deal structuring, negotiating term sheets, performing due diligence;Entity organization, corporate structuring and corporate governance; andSupporting the transactional department. The ideal candidate will have a proven track record of delivering superior results as well as the following skills and experience:Demonstrated ability to work well under pressure and manage multiple critical projects and deadlines;Strong business judgment and strategic thinking;Superior attention to detail;Ability to independently manage and review projects with minimal partner oversight;Proactive, resourceful, entrepreneurial approach to recognizing needs and solving problems;A team-player who is result-oriented and self-motivated;Able to grasp new concepts quickly and efficiently; andExcellent verbal and written communication skills. The position includes a competitive base salary starting at $150,000 plus a comprehensive benefits package. To apply, please submit a resume and your salary requirements.

## Output
{{
    tasks: ["Drafting, reviewing and negotiating various corporate and commercial agreements", "Deal structuring, negotiating term sheets, performing due diligence", "Entity organization, corporate structuring and corporate governance",  "Supporting the transactional department."]
}}

# Input
{description}

# Output
Return the task descriptions in the following json format:
{{
    tasks: [List of all tasks in the job description]
}}
\end{lstlisting}
\caption{Prompt used to extract the job description from a given posting. Fields in braces are filled in for each posting.}
\label{fig:description-extraction-prompt}
\end{figure*}

\begin{figure*}[t]
\begin{lstlisting}[basicstyle=\ttfamily\scriptsize, breaklines=true, columns=fullflexible, frame=single, framesep=4pt, xleftmargin=4pt, xrightmargin=4pt]
Your job is to decompose the job task into a list of tasks written at the same granularity of an O*Net IWA. Ensure that it matches the level of generality of an O*Net IWA: it should be general enough to be an activity performed in a large number of occupations across multiple job families, but specific enough to capture the essence of the task.

# Task Details
1. Evaluate the job task and identify whether it is at the same granularity of an O*Net IWA or at a less granular level.
2. If the job task is at the same granularity of an O*Net IWA, return the job task in the style of an O*Net IWA.
3. If the job task is at a less granular level, decompose the job task into a list of tasks written in the styleof an O*Net IWA. Tasks should be unique and not redundant.
4. Return the list of tasks in the following JSON format. If the job task is at the same granularity of an O*Net IWA, there should be only one task.
{{
    "tasks": [
        "Task 1",
        "Task 2",
        "Task 3",
        ...
    ]
}}
5. Try to decompose the job tasks into the fewest number of tasks possible while maintaining the same granularity of an O*Net IWA.

## Examples of IWAs
- Coordinate regulatory compliance activities.
- Direct security or safety activities or operations.
- Repair tools or equipment.
- Assist scientists, scholars, or technical specialists with projects or research.
- Communicate with others about operational plans or activities.
- Coordinate with others to resolve problems.

# Example
Input:
Job Task: "Handle all of the financial activities at the firm."
Output:
{{
    "tasks": [
        "Manage financial activities",
    ]
}}

Input:
Inspect, clean, test, and maintain respiratory therapy equipment to ensure equipment is functioning safely and efficiently and notify manager/supervisor when repairs are necessary.
Output:
{{
    "tasks": [
        "Inspect respiratory therapy equipment",
        "Clean respiratory therapy equipment",
        "Test respiratory therapy equipment",
        "Maintain respiratory therapy equipment",
        "Notify manager/supervisor when repairs are necessary",
        "Ensure equipment is functioning safely and efficiently",
    ]
}}

# Input
Job Task: {params['task']}

# Output
Return the list of tasks in the following JSON format:
{{
    "tasks": [
        "Task 1",
        "Task 2",
        "Task 3",
        ...
    ]
}}
\end{lstlisting}
\caption{Prompt used to standardize the extracted tasks from a job posting. Fields in braces are filled in for each posting.}
\label{fig:task-standardize-prompt}
\end{figure*}

\begin{figure*}[t]
\begin{lstlisting}[basicstyle=\ttfamily\scriptsize, breaklines=true, columns=fullflexible, frame=single, framesep=4pt, xleftmargin=4pt, xrightmargin=4pt]
You are checking which work activities are described in a job posting.

Job title: {title}

Job description:
{posting text}

Below is a list of work activities. For each one, decide whether the job description says or clearly implies that a \
person in this job performs this activity as part of their work.

Count an activity as present only if:
- the posting describes duties that are an instance of this activity (wording can differ; for example, "ring up \
purchases" is an instance of "Execute financial transactions"), and
- the worker in this job performs it, not the company, a customer, or another team.
Do not count an activity if it is only loosely related to the job, appears only as a required skill, degree, or \
qualification without a corresponding duty, or appears only in company boilerplate.

Activities:
{items}

Return JSON covering every activity:
{{"judgments": [{{"id": "A1", "present": true, "evidence": "short quote from the posting, or empty"}}, ...]}}"""
\end{lstlisting}
\caption{Prompt used to verify the work activities with content from the original job posting. Fields in braces are filled in for each posting.}
\label{fig:verifier-prompt}
\end{figure*}

\section{Measuring AI Mentions in Job Postings}
\label{app:ai-mentions}

This appendix describes how we identify and classify mentions of AI in job postings (Sec.~\ref{sec:visible}) and reports additional results. Unless noted otherwise, all estimates use the 39,578 postings from 33 firms posted between 2023 and 2025.

\subsection{Why We Use Raw Job Descriptions}
\label{app:ai-mentions-raw}

The IWA representation used in the rest of the paper is built from tasks that are extracted from each posting and then standardized to the O*NET taxonomy (Sec.~4.1). This process retains what a role does but largely discards how the posting refers to AI. Among postings whose description mentions AI, an AI-related term (Table~\ref{tab:ai-keywords}) appears in the extracted tasks of 46.3\% and in the decomposed tasks of 37.5\%. Among postings that expect workers to use AI tools, the corresponding shares are 63.4\% and 54.0\%. We therefore measure AI mentions from the full text of each job description.

\subsection{Implementation}
\label{app:ai-mentions-keywords}

\paragraph{Keyword Matching.} Table~\ref{tab:ai-keywords} lists the terms we match. We match terms on word boundaries. Acronyms (AI, A.I., ML, NLP, LLM, GPT) are matched case-sensitively so that they do not match ordinary words, and all other terms are matched case-insensitively. Before matching, we remove URLs, e-mail addresses, and web domains ending in \texttt{.ai} (e.g., career-site domains), which otherwise produce spurious matches. We then split each description into sentences at line breaks and sentence-final punctuation and remove list markers. A posting is flagged if any sentence contains a term from the \textit{Any AI} tier. We report the \textit{Generative AI} and \textit{AI tools} tiers as narrower keyword measures (Table~\ref{tab:ai-mentions-trends}).

\begin{table}[t]
  \centering
  \small
  \begin{tabular}{@{}p{0.22\columnwidth}p{0.70\columnwidth}@{}}
    \toprule
    Tier & Terms \\
    \midrule
    Any AI & AI, A.I., ML, NLP, artificial intelligence, machine learning, deep learning, neural network(s), natural language processing, computer vision, reinforcement learning, and all terms below \\
    \addlinespace
    Generative AI & LLM(s), GPT (incl.\ versions, e.g., GPT-4), AI agent(s), generative AI, GenAI, large language model(s), foundation model(s), ChatGPT, OpenAI, Copilot, prompt engineering, agentic, retrieval-augmented generation \\
    \addlinespace
    AI tools & AI tool(s), AI assistant(s), AI copilot(s) (incl.\ ``AI-powered/-based/-enabled/-driven'' variants and GenAI), AI-assisted, Copilot, ChatGPT, generative AI tool(s), generative AI assistant(s) \\
    \bottomrule
  \end{tabular}
  \caption{AI-related terms used to flag job postings. Tiers are nested: a posting that matches the \textit{Generative AI} or \textit{AI tools} tier also matches \textit{Any AI}. Acronyms are matched case-sensitively.}
  \label{tab:ai-keywords}
\end{table}

\paragraph{Removing Firm Boilerplate.}
\label{app:ai-mentions-boilerplate}

Firms often append the same AI-related language to postings for unrelated roles. For example, the sentence ``We're Salesforce, the Customer Company, inspiring the future of business with AI+ Data +CRM'' appears in 505 Salesforce postings across 456 distinct job titles. To avoid counting such text as a mention of AI in the role, we normalize each AI-related sentence (lowercasing, replacing digits, and removing punctuation) and flag it as boilerplate if, within the same firm and half-year, it (i)~appears in at least 10\% of the firm's postings, (ii)~appears under at least five distinct job titles, and (iii)~contains at least eight words. Condition (ii) prevents re-posted openings for the same role from being treated as boilerplate, and condition (iii) keeps short, generic requirements (e.g., a one-line request for AI/ML experience) from being removed. This rule flags 1,855 of the 17,802 AI-related sentences in the dataset. Across half-years between 2023 and 2025, 6--25\% of postings with any AI-related term contain only boilerplate AI language. Results are not sensitive to the 10\% threshold (Table~\ref{tab:ai-mentions-robustness}).

\paragraph{Classifying How Postings Refer to AI.}
\label{app:ai-mentions-llm}

For each posting with at least one non-boilerplate AI-related sentence, we prompt an LLM (gpt-6-luna, version 2026-09-22) with the job title, the firm name, and up to 12 of the posting's AI-related sentences (each truncated to 400 characters). The model assigns every applicable label from Table~\ref{tab:ai-mention-labels} and returns a short supporting quote. The full prompt is shown in Fig.~\ref{fig:ai-mention-prompt}. We use the model's default decoding settings and request JSON output; all 4,006 calls returned a valid response.

We count a posting as mentioning AI if it contains at least one non-boilerplate AI-related term and the model does not label it \textsc{Not\_AI} only. This removes 26 of the 2,887 keyword-matched postings between 2023 and 2025 (e.g., ``Tax LLM preferred,'' which refers to a law degree, and ``Proficient in Adobe AI \& PS''), leaving 2,861 AI-mentioning postings.

Labels are not mutually exclusive. Of the 2,861 AI-mentioning postings, 2,097 (73.3\%) receive one label, 666 (23.3\%) receive two, and 98 (3.4\%) receive three or more. Table~\ref{tab:ai-mention-cooccurrence} reports how often labels co-occur; the most common pair is \textit{Builds AI} and \textit{AI knowledge required}. 

\textit{Uses AI tools} captures whether the role uses AI in its own work, not whether the posting names a tool. Of the 276 postings with this label between 2023 and 2025, 52 (18.8\%) name a specific product such as Copilot or ChatGPT, and 100 (36.2\%) match the \textit{AI tools} keyword tier (Table~\ref{tab:ai-keywords}). The remainder describe applying AI to the worker's own tasks without naming a tool (e.g., ``exploring opportunities to utilize AI for monitoring'' in a compliance manager posting). Conversely, only 100 of the 297 postings that match the \textit{AI tools} keyword tier (33.7\%) receive this label. Most of the others describe building, supporting, or knowing about AI tools rather than using them (e.g., a research internship on Copilot tuning, labeled \textit{Builds AI}).

\begin{table*}[t]
  \centering
  \small
  \begin{tabular}{@{}p{0.15\textwidth}p{0.36\textwidth}p{0.43\textwidth}@{}}
    \toprule
    Label & Definition given to the model & Example (job title: supporting quote) \\
    \midrule
    Builds AI & The role designs, develops, trains, fine-tunes, evaluates, or deploys AI/ML models or AI systems. & UPS Capital Parcel Pro AI Summer 2024 Intern: ``Design, develop and test AI solutions critical to customer experience and decision making.'' \\
    \addlinespace
    Uses AI tools & The role is expected to use AI tools or assistants in its own work, or to apply AI to improve its own workflows. & Sr.\ Legal Specialist -- Practice Management: ``Skilled in AI tools (e.g., Copilot, CLM systems, automation platforms)'' \\
    \addlinespace
    AI product or team & The role sells, markets, manages, supports, secures, or governs AI products or features, or sits in an AI-focused team, without itself building the models. & Strategy \& Planning Manager Sr (Intelligent Automation): ``own the Virtual Assistant (chatbot) strategy, develop and support implementation of use cases'' \\
    \addlinespace
    AI knowledge required & The posting asks for knowledge of, familiarity with, or interest in AI/ML as a qualification, without saying the role builds or uses it in daily work. & Director, Technical Program Management (Customer Success): ``A strong understanding of AI and Agentic concepts'' \\
    \addlinespace
    General context only & AI appears only as company, team, industry, or benefits context. & Data Engineering II: ``The Cloud \& AI organization accelerates Microsoft's mission'' \\
    \addlinespace
    Not AI & The matched terms do not refer to artificial intelligence. & Summer Intern, Apparel Design: ``Proficient in Adobe AI \& PS'' \\
    \bottomrule
  \end{tabular}
  \caption{Labels used to classify how a job posting refers to AI. A posting can receive more than one label, except that \textit{General context only} is used only when no other AI label applies and \textit{Not AI} only when no sentence refers to AI. Examples are drawn from postings that received a single label.}
  \label{tab:ai-mention-labels}
\end{table*}

\begin{table}[t]
  \centering
  \small
  \setlength{\tabcolsep}{4pt}
  \begin{tabular}{@{}lrrrr@{}}
    \toprule
     & Build & Use & Prod. & Know. \\
    \midrule
    Builds AI & \textbf{1,139} & 45 & 121 & 400 \\
    Uses AI tools & 45 & \textbf{276} & 30 & 70 \\
    AI product/team & 121 & 30 & \textbf{634} & 303 \\
    AI knowledge & 400 & 70 & 303 & \textbf{1,429} \\
    \bottomrule
  \end{tabular}
  \caption{Co-occurrence of labels among the 2,861 AI-mentioning postings from 2023--2025. Diagonal cells (bold) give the number of postings with each label; off-diagonal cells give the number with both labels. Columns follow the row order (Build = \textit{Builds AI}, Use = \textit{Uses AI tools}, Prod.\ = \textit{AI product or team}, Know.\ = \textit{AI knowledge required}). A further 248 postings are labeled \textit{General context only}, which by construction does not co-occur with other labels.}
  \label{tab:ai-mention-cooccurrence}
\end{table}

\begin{figure*}[t]
\begin{lstlisting}[basicstyle=\ttfamily\scriptsize, breaklines=true, columns=fullflexible, frame=single, framesep=4pt, xleftmargin=4pt, xrightmargin=4pt]
You are coding how a job posting talks about artificial intelligence (AI).

Job title: {title}
Company: {firm}
Sentences from the posting that contain AI-related terms (company-wide boilerplate already removed):
{numbered AI-related sentences}

Assign EVERY label that applies to how THIS ROLE relates to AI:
- BUILD_AI: the role designs, develops, trains, fine-tunes, evaluates, or deploys AI/ML models or AI systems (including LLM applications, AI agents, ML pipelines).
- USE_AI_TOOLS: the role is expected to use AI tools or assistants in its own work (e.g., GitHub/Microsoft Copilot, ChatGPT, generative-AI tools, AI-powered software), or to apply AI to improve its own workflows.
- AI_PRODUCT_OR_TEAM: the role sells, markets, manages, supports, secures, or governs AI products or features, or sits in an AI-focused team, without itself building the models.
- AI_KNOWLEDGE: the posting asks for knowledge of, familiarity with, or interest in AI/ML as a qualification, without saying the role builds or uses it in daily work.
- GENERAL_CONTEXT: AI appears only as company, team, industry, or benefits context, not as part of this role's work or requirements.
- NOT_AI: the matched terms do not refer to artificial intelligence (e.g., "AI" meaning avian influenza, Adobe Illustrator, artificial insemination, or action item).

Rules: use NOT_AI only if none of the sentences refer to artificial intelligence. Use GENERAL_CONTEXT only if no other AI label applies. Judge only from the text given.

Respond with JSON only: {"labels": [<labels>], "evidence": "<quote of at most 20 words supporting the most specific label>"}
\end{lstlisting}
\caption{Prompt used to classify how each job posting refers to AI. Fields in braces are filled in for each posting.}
\label{fig:ai-mention-prompt}
\end{figure*}

\subsection{Estimation}
\label{app:ai-mentions-estimation}

To estimate the change in the share of postings that mention AI, we fit a linear probability model at the posting level with fixed effects for firm and SOC major group:
\begin{equation}
  \texttt{AI Mention} \sim \texttt{Year} + \texttt{Firm} + \texttt{SOC}
\end{equation}
\texttt{Year} is the number of years between January 1, 2023, and the posting date, so its coefficient gives the annual change in the probability of an AI mention, in percentage points. We compute cluster-robust standard errors by firm (CR1) and derive confidence intervals and $p$-values from a $t$ distribution with $G-1$ degrees of freedom, where $G = 33$ firms. We fit the same model separately for each label and keyword tier. For firm-level comparisons, we compare the share of AI-mentioning postings in 2023 and 2025 for each of the 20 firms with at least 200 postings in both years using a two-proportion $z$-test and apply the Benjamini-Hochberg procedure across firms.

\begin{table*}[t]
  \centering
  \small
  \begin{tabular}{@{}lrrrrcl@{}}
    \toprule
    & \multicolumn{3}{c}{\% of postings} & \multicolumn{3}{c}{Annual change (pp)} \\
    \cmidrule(lr){2-4} \cmidrule(l){5-7}
    Measure & 2023 & 2024 & 2025 & $\hat\beta$ & 95\% CI & $p$ \\
    \midrule
    AI mention (main measure) & 5.28 & 7.33 & 8.76 & 2.37 & [0.58, 4.16] & .011 \\
    \quad excluding \textit{General context only} & 4.86 & 6.56 & 8.10 & 2.22 & [0.54, 3.90] & .011 \\
    \addlinespace
    \multicolumn{7}{@{}l}{\textit{LLM labels (not mutually exclusive)}} \\
    \quad Builds AI & 2.19 & 2.78 & 3.54 & 1.17 & [$-$0.04, 2.37] & .057 \\
    \quad Uses AI tools & 0.16 & 0.34 & 1.49 & 0.69 & [0.35, 1.02] & $<$.001 \\
    \quad AI product or team & 0.97 & 1.50 & 2.22 & 0.64 & [0.11, 1.17] & .019 \\
    \quad AI knowledge required & 2.78 & 3.79 & 4.14 & 0.94 & [0.21, 1.67] & .013 \\
    \quad General context only & 0.42 & 0.78 & 0.66 & 0.15 & [$-$0.04, 0.34] & .120 \\
    \addlinespace
    \multicolumn{7}{@{}l}{\textit{Keyword tiers (boilerplate removed, no LLM filter)}} \\
    \quad Any AI & 5.36 & 7.39 & 8.82 & 2.36 & [0.57, 4.15] & .011 \\
    \quad Generative AI & 0.93 & 2.12 & 3.44 & 1.58 & [0.26, 2.90] & .020 \\
    \quad AI tools & 0.20 & 0.54 & 1.40 & 0.67 & [0.17, 1.17] & .010 \\
    \midrule
    Postings & 11,902 & 13,419 & 14,257 & \multicolumn{3}{l}{33 firms in each year} \\
    \bottomrule
  \end{tabular}
  \caption{Share of job postings mentioning AI, by measure and year, and the estimated annual change from a linear probability model with firm and SOC major group fixed effects (standard errors clustered by firm).}
  \label{tab:ai-mentions-trends}
\end{table*}

\begin{table*}[t]
  \centering
  \small
  \begin{tabular}{@{}lrrrrcrr@{}}
    \toprule
    & \multicolumn{2}{c}{\% of postings} & \multicolumn{3}{c}{Annual change (pp)} & & \\
    \cmidrule(lr){2-3} \cmidrule(lr){4-6}
    Specification & 2023 & 2025 & $\hat\beta$ & 95\% CI & $p$ & Postings & Firms \\
    \midrule
    Main specification & 5.28 & 8.76 & 2.37 & [0.58, 4.16] & .011 & 39,578 & 33 \\
    No fixed effects & 5.28 & 8.76 & 2.27 & [0.22, 4.33] & .031 & 39,578 & 33 \\
    Excluding \textit{General context only} & 4.86 & 8.10 & 2.22 & [0.54, 3.90] & .011 & 39,578 & 33 \\
    \addlinespace
    \multicolumn{8}{@{}l}{\textit{Keyword measure (Any AI tier, no LLM filter)}} \\
    \quad Boilerplate threshold 10\% (default) & 5.36 & 8.82 & 2.36 & [0.57, 4.15] & .011 & 39,578 & 33 \\
    \quad Boilerplate threshold 5\% & 5.36 & 8.82 & 2.41 & [0.61, 4.20] & .010 & 39,578 & 33 \\
    \quad Boilerplate threshold 20\% & 5.51 & 8.89 & 2.33 & [0.56, 4.11] & .012 & 39,578 & 33 \\
    \quad No boilerplate removal & 6.64 & 9.71 & 2.40 & [0.57, 4.22] & .012 & 39,578 & 33 \\
    \addlinespace
    \multicolumn{8}{@{}l}{\textit{Sample restrictions}} \\
    \quad Excluding Salesforce and Microsoft & 3.60 & 5.40 & 1.32 & [0.30, 2.34] & .013 & 36,961 & 31 \\
    \quad Excluding Information sector & 3.50 & 5.37 & 1.30 & [0.21, 2.39] & .021 & 35,205 & 29 \\
    \quad Excluding Computer \& Math.\ occupations & 2.25 & 4.54 & 1.30 & [0.31, 2.30] & .012 & 33,181 & 33 \\
    \bottomrule
  \end{tabular}
  \caption{Robustness of the estimated annual change in the share of postings mentioning AI. Unless noted, estimates use the main AI-mention measure and a linear probability model with firm and SOC major group fixed effects, with standard errors clustered by firm.}
  \label{tab:ai-mentions-robustness}
\end{table*}

\begin{table}[t]
  \centering
  \small
  \setlength{\tabcolsep}{5pt}
  \begin{tabular}{@{}lrrrr@{}}
    \toprule
    Firm & 2023 & 2025 & $\Delta$ (pp) & $q$ \\
    \midrule
    Salesforce & 11.7 & 43.1 & $+$31.4 & $<$.001 \\
    Microsoft & 38.6 & 68.8 & $+$30.2 & $<$.001 \\
    General Motors & 6.6 & 18.3 & $+$11.7 & $<$.001 \\
    Comcast & 6.5 & 9.7 & $+$3.2 & .275 \\
    McKesson & 8.2 & 10.1 & $+$1.9 & .424 \\
    ExxonMobil & 2.1 & 3.8 & $+$1.7 & .424 \\
    Morgan Stanley & 4.7 & 6.3 & $+$1.6 & .644 \\
    Five Below & 0.0 & 0.8 & $+$0.8 & .289 \\
    3M & 3.6 & 4.3 & $+$0.7 & .808 \\
    Corteva Agriscience & 7.8 & 8.5 & $+$0.7 & .808 \\
    Fluor Corporation & 0.0 & 0.3 & $+$0.3 & .424 \\
    Walmart & 8.1 & 8.3 & $+$0.2 & .919 \\
    Cleveland-Cliffs & 0.5 & 0.5 & $+$0.0 & .960 \\
    Duke Energy & 1.4 & 1.3 & $-$0.1 & .919 \\
    Xcel Energy & 1.9 & 1.6 & $-$0.3 & .808 \\
    Freeport-McMoRan & 1.6 & 1.1 & $-$0.5 & .557 \\
    UPS & 1.5 & 0.9 & $-$0.6 & .424 \\
    Primoris Services & 1.9 & 1.0 & $-$0.9 & .424 \\
    Wesco & 1.5 & 0.0 & $-$1.5 & .037 \\
    Bank of America & 6.5 & 2.7 & $-$3.8 & .037 \\
    \bottomrule
  \end{tabular}
  \caption{Share of each firm's job postings mentioning AI in 2023 and 2025, for the 20 firms with at least 200 postings in both years. $q$-values are Benjamini-Hochberg adjusted $p$-values from two-proportion $z$-tests.}
  \label{tab:ai-mentions-firms}
\end{table}

\begin{table}[t]
  \centering
  \small
  \begin{tabular}{@{}lrr@{}}
    \toprule
    SOC major group & 2023 & 2025 \\
    \midrule
    Computer \& Mathematical & 18.8 & 36.3 \\
    Management & 5.8 & 11.8 \\
    Life, Physical \& Social Science & 10.4 & 8.8 \\
    Arts, Design, Entertainment \& Media & 2.2 & 6.2 \\
    Architecture \& Engineering & 2.2 & 4.6 \\
    Business \& Financial Operations & 1.9 & 4.5 \\
    Office \& Administrative Support & 0.1 & 1.2 \\
    Sales \& Related & 0.6 & 1.1 \\
    Production & 0.0 & 0.3 \\
    Construction \& Extraction & 0.0 & 0.0 \\
    Installation, Maintenance \& Repair & 0.0 & 0.0 \\
    Transportation \& Material Moving & 0.0 & 0.0 \\
    \bottomrule
  \end{tabular}
  \caption{Share of job postings mentioning AI (\%) by SOC major group, for groups with at least 300 postings between 2023 and 2025, ordered by the 2025 share.}
  \label{tab:ai-mentions-soc}
\end{table}

\section{Analyzing Change in Job Postings}
\label{app:taskcount}

In Sec.~\ref{sec:prev}, we discuss how job postings in our dataset have fewer work activities overall, potentially contributing to the observed association between how AI-exposed a work activity is and its declining prevalence. We conduct further analyses to understand where this decline originates. These analyses use the 22,332 postings from the 20 firms with at least 200 postings in both 2023 and 2025.

\subsection{Methods}
Our pipeline extracts responsibilities from each job description, decomposes them into tasks, matches each task to its most similar O*NET task, maps O*NET tasks to IWAs, and verifies each IWA against the job description. Holding the firm and normalized job title fixed, we decompose the change in the number of verified IWAs per posting into the change at each of these stages, so that the log change at each stage sums approximately to the total log change. To check that the decline is not due to variability across extraction runs, we also re-extracted responsibilities from 2023 and 2025 postings on the same day with the same model.

\paragraph{Identifying firm templates.} Many firms reuse blocks of text across postings. We use the term ``template responsibilities'' to refer to an extracted line that appears in postings for at least 10 distinct job titles and in at least 2\% of a firm's postings in a given year, and measure how much of the decline in extracted responsibilities comes from template lines.

\paragraph{Conducting a matched-pair audit.} To examine how the text of job descriptions changed, we also matched 2023 and 2025 postings
from the same firm with the same normalized title and seniority. We drew two samples: 50 randomly selected pairs, and 25 pairs from 13 firms in which the 2025 posting yields at most 70\% as many extracted responsibilities as the 2023 posting. We then compare the paired postings using the following three steps. First, reading only the raw job description, we prompt an LLM (GPT-6 Luna) to list each distinct duty the worker performs, excluding qualifications, values, company descriptions,benefits, and legal text, and estimated the share of the text devoted to each type of content. Second, the model compared the two postings in each pair, presented in random order without dates, and judged which lists more duties, which is
worded more generically, and which contains more non-duty text. Third, it compared its list of duties with the
responsibilities extracted by our pipeline to identify duties the pipeline missed. One author read approximately ten
pairs in full and found that the audited duty counts and missed duties were consistent with the raw text.

\subsection{Findings}                                              \paragraph{The decline in work activities begins at extraction.} 
The decline in verified IWAs arises at the first stage of the pipeline --- extracting responsibilities from the job postings. Extracted responsibilities fell by 13.2 log points within firm and title, while decomposition, O*NET matching, and    verification together offset part of this decline, for a total decline of 9.6 log points in verified IWAs. We find no evidence that newer postings contain more redundant responsibilities that collapse into fewer IWAs.      

\paragraph{Firm templates account for most of the decline.} Template responsibilities fell from 2.0 to 1.1 per
posting, accounting for 58\% of the overall decline in extracted responsibilities. Much of this comes from a single firm, Five Below. For Five Below, the template listed approximately 30 duties in most 2023 postings and only 6 in 2025, accounting  for roughly 40\% of the decline. Excluding Five Below, extracted responsibilities still decline by 8.2 log points. Some 2023 template lines were values or behavioral statements (e.g., ``Win as a team'') rather than duties, while
others described real duties that later postings condensed or removed.

\paragraph{Newer postings list fewer and broader duties.} In the 25 pairs with large declines, the number of duties   identified directly from the job description fell from 26.7 to 16.9 ($p = .001$), and the share of text devoted to duties fell from 32\% to 21\% ($p = .003$), while the share devoted to pay and benefits rose from 12.1\% to 15.7\%. Among these pairs, 64\% had at least 15\% fewer written duties, 64\% were judged to contain more non-duty text in 2025, and 60\% were judged to be worded more generically in 2025. These explanations overlap. In the 50 random pairs, neither extracted responsibilities (9.96 to 9.30, $p = .51$) nor audited duties (19.2 to 18.6, $p = .67$) declined significantly; with 50 pairs, the audit is underpowered to detect a population-wide change.

\paragraph{Extraction also contributes to the decline.} The share of written duties that the pipeline missed rose from 7.0\% to 13.7\% in the large-decline pairs ($p = .05$) and from 8.7\% to 12.5\% in the random pairs ($p = .19$). The extractor also varies in granularity, sometimes splitting a list of duties into several tasks and sometimes keeping it as one.  Changes in the job descriptions therefore explain much, but not all, of the decline.                                   Table~\ref{tab:jd_examples} presents representative examples of each type of change.

\begin{table*}[t]
\centering
\small
\begin{tabularx}{\linewidth}{p{2.8cm} cc X}
\toprule
Firm, role & Extracted & Duties & What changed \\
\midrule
\multicolumn{4}{l}{\emph{Firm template rewrite}} \\
\midrule
Five Below, Merchandise Manager & 32 $\to$ 8 & -- & About 30 one-line duties (``Unloads merchandise from trucks,'' ``Checks in shipments,'' ``Stages merchandise for the sales floor'') are condensed into eight broader bullets (``Oversees the receiving, unpacking, and organizing of new stock''). \\
Walmart, Developmental Technician HVAC/R & 17 $\to$ 5 & 27 $\to$ 13 & Seven duty paragraphs and competency statements (``Models the Walmart values'') are replaced by recruiting text and a five-bullet duty list. \\
\addlinespace
\midrule
\multicolumn{4}{l}{\emph{Fewer and broader duties}} \\
\midrule
McKesson, Pricing Analyst & 10 $\to$ 3 & 23 $\to$ 6 & Ten specific duties (SQL and Power BI tools, price benchmarking, performance monitoring) are summarized as three broad duties (``Perform customer level pricing analysis for regional customers''). \\
UPS, Marketing Intern & 19 $\to$ 4 & 34 $\to$ 12 & A duty list repeated in two forms is replaced by four abstract research duties. \\
\addlinespace
\midrule
\multicolumn{4}{l}{\emph{Fewer duties, more non-duty text}} \\
\midrule
Fluor, Contract Admin Spec III & 20 $\to$ 9 & 43 $\to$ 22 & The second duty list (contract negotiation, RFP meetings, closeout) is removed and two paragraphs on the company's purpose are added. \\
\addlinespace
\midrule
\multicolumn{4}{l}{\emph{Extraction inconsistency}} \\
\midrule
Wesco, Senior Global Account Manager & 21 $\to$ 11 & 33 $\to$ 31 & Nearly identical duty text, but the 2025 extraction omits ten duties (e.g., ``maintain executive level client relationships''). \\
Duke Energy, Commodities Services Technician & 13 $\to$ 4 & 10 $\to$ 11 & Nearly identical duty text; a list of duties is split into separate tasks in 2023 and kept as one task in 2025. \\
\bottomrule
\end{tabularx}
\caption{We provide representative matched pairs of 2023 and 2025 postings from the same firm and job title. For each pair, we list (1)~Extracted: the number of responsibilities extracted by our pipeline; (2)~Duties: is the number of distinct duties identified directly from the job description in the audit; (3)~Description qualitatively capturing what changed from 2023 to 2025.}
\label{tab:jd_examples}
\end{table*}

\section{AI-Linked Responsibilities Within Existing Activities}\
\label{app:change_analysis}
\subsection{Identifying AI-Linked Work Activities}
\label{app:ai_linkage}

The analysis in Sec.~7 is conducted on posting--IWA pairs: each verified IWA in a posting is counted once, regardless of how many tasks support it. To decide whether a pair is AI-linked, we trace the non-boilerplate AI-related sentences of each job description (Appendices~\ref{app:ai-mentions-keywords}) through both pipeline runs. An extracted task
is AI-related if it contains an AI-related term or reproduces an AI-related sentence (it covers at least half of the sentence's content words, or at least 70\% of its own content words appear in the sentence). The pipeline does not store which decomposed tasks come from which extracted task, so we assign decomposed tasks to their parent with an order-preserving lexical alignment; this alignment agrees with an LLM's assignment for 93\% of decomposed
tasks in 40 audited postings. Each decomposed task carries its matched O*NET task and the IWAs that task maps to. A verified IWA is AI-linked if (i) the verifier's evidence quote for it contains an AI-related term, or (ii) it is reached from an AI-related task and at least half of the words of the verifier's evidence quote appear in that task or in its source sentence. To restrict the measure to AI-related \emph{responsibilities}, we locate the
supporting AI text in the description, using the description's section headers (a sentence or quote is placed in
a section when at least 80\% of its words appear in one of its lines). We drop links whose supporting text lies
in qualification sections or in organizational context (company and values statements, pay and benefits, legal and
logistical text), and retain text in responsibility or role-summary sections, as well as text that cannot be
located. Firm boilerplate (Appendix~\ref{app:ai-mentions-keywords}) is excluded before tracing. This filter removes 97 of 1{,}547 AI-linked
pairs (6.3\%). For IWA $i$ in year $t$, the AI-Linked Share is
\begin{equation}
  s_{i,t} = \frac{\bigl|\{p \in P_t : i \in V_p \text{ and } (p,i) \text{ is AI-linked}\}\bigr|}
                 {\bigl|\{p \in P_t : i \in V_p\}\bigr|},
\end{equation}
where $P_t$ is the set of postings in year $t$ and $V_p$ is the set of verified IWAs of posting $p$.

AI-linked denotes AI-related content within an activity, not the use of AI to perform it. AI can be the object
of the work rather than a tool: ``develop, train, test, and deploy AI tools'' supports \emph{Design computer or
information systems or applications}. The verifier establishes that an IWA is supported by the job description and
quotes the supporting text; it does not validate each extracted-to-decomposed-to-O*NET mapping. Task-level provenance
is therefore approximate, because it relies on a heuristic alignment and on nearest-neighbor O*NET matches. The flag
should be read as evidence that a verified activity is described with AI-related content in the posting, not as a
validated mapping of an individual AI task to that activity. Because AI-related text is identified by keywords
(Table~5), AI-related work described without such terms is not counted.

\subsection{Change in AI-Linked Share versus Prevalence}
In Sec.~\ref{sec:content}, we discuss how AI-related responsibilities can increase in a work activity even as the overall prevalence declines. In Table~\ref{tab:ai_share_declining}, we list the 15 IWAs that most frequently carry AI-related content. These work activities account for 62\% of AI-linked posting--IWA pairs. Fourteen of the 15 decline in overall presence between 2023 and 2025, while 14 of the 15 show an increase in AI-Linked Share. Across all activities, AI-linked activity remains a small share of the total, growing from 0.6\% to 1.3\% of verified activity between 2023 and 2025.

\paragraph{Mapping reliability.} Rankings of individual IWAs should be interpreted with caution. In an audit of 1,000 candidate links between AI-related tasks and IWAs, rated by two independent LLM raters ($\kappa \approx 0.66$--$0.71$) with manual review of a subset, most links the verifier did not confirm reflected errors in decomposition or O*NET matching rather than work outside the taxonomy, and even confirmed links were frequently judged to be imperfect matches. The best-fitting existing IWA usually captures the general function of an AI-related responsibility (e.g., designing or programming a system) but omits the AI-specific object or method in 69\% of cases. Two IWAs in Table~\ref{tab:ai_share_declining}, marked with $\dagger$, are largely mapping artifacts: for example, ``develop machine learning models for recommendation systems'' was matched to ``advise others on business or operational matters'' through an O*NET task about recommending data-driven solutions. Correcting the links changes the rankings of individual IWAs but not the overall concentration of AI-related work in highly exposed activities (54\% vs.\ 51\% of audited links in the top Copilot exposure quartile).
\input{Figures/Table/IWAShare}

\section{Robustness Analysis}
\label{app:robustness}
\subsection{Using Different Definitions of Change in Frequency}
In Sec.~\ref{sec:prev}, we define change in frequency as the difference in the percentage of postings that mention the work activity between 2023 and 2025. As shown in Table~\ref{tab:robustness}, we evaluate our pooled model using four additional operationalizations of ``change in frequency,'' fit using our two measures of AI Exposure. We observe directional consistency across all of our operationalizations, supporting the negative association between how AI-exposed a work activity is and the change in frequency from 2023 to 2025. 

\begin{table*}[t]
\centering
\small
\begin{tabular}{p{2.5cm} p{6.5cm} cc}
\toprule
{Outcome} & {Definition}
& {Copilot Usage} & {Claude Usage} \\
\midrule

Conditional binomial
& Log rate ratio of activity mentions in 2025 versus 2023, adjusting for posting volume
& $-0.028^{*}$ & $-0.042^{**}$ \\
& & $(0.013)$ & $(0.015)$ \\[0.3em]

Mean IWA strength
& Change in average normalized IWA weight
& $-0.010$ & $-0.031$ \\
& & $(0.009)$ & $(0.016)$ \\[0.3em]

Percentile rank
& Change in an activity's within-firm percentile rank by presence
& $-0.650^{**}$ & $-0.542$ \\
& & $(0.249)$ & $(0.303)$ \\[0.3em]

Scaled change
& Change in presence divided by mean presence, restricted to activities with mean presence $\geq 1\%$
& $-3.798$ & $-3.787$ \\
& & $(2.819)$ & $(2.162)$ \\[0.3em]

\bottomrule
\end{tabular}
\caption{Robustness of the association between AI exposure and changes
in advertised work activity frequency. Coefficients represent the
association with a one-standard-deviation increase in log AI exposure.
Standard errors in parentheses. All models use log-transformed, standardized AI exposure and exclude activities with exposure $\leq 10^{-6}$. $^{*}p<0.05$, $^{**}p<0.01$, $^{***}p<0.001$.
}
\label{tab:robustness}
\end{table*}

\subsection{Applying Different Posting Thresholds}
\input{Figures/FloorComp}
In Sec.~\ref{sec:prev}, we filter for firms that have at least 200 postings in each time period. We report results for alternate thresholds in Table~\ref{tab:floor_robustness}. Across all definitions of our dependent variable and all thresholds, we observe a negative relationship between AI exposure and the change in prevalence.

\subsection{Accounting for Firm-Specific Behaviors}
\label{app:ai-mentions-results}
Table~\ref{tab:ai-mentions-trends} reports annual shares and estimated trends for each measure. Table~\ref{tab:ai-mentions-robustness} reports the main estimate under alternative specifications. Salesforce and Microsoft together account for 42.1\% of AI-mentioning postings in 2025, so we also report estimates without them. The increase is robust to removing the LLM filter, to omitting or varying the boilerplate rule, and to excluding Salesforce and Microsoft, the Information sector, or Computer \& Mathematical occupations, although its magnitude roughly halves when the firms or occupations with the highest rates of AI mentions are excluded. Tbls.~\ref{tab:ai-mentions-firms} and~\ref{tab:ai-mentions-soc} report shares by firm and by occupation group.

%% file: Figures/JobTitleExamples.tex
\begin{table}
    \centering
    \small
    \begin{tabular}{@{}p{0.43\linewidth}p{0.43\linewidth}}
    \toprule
    Original Title &  O*NET Title \\
    \midrule
    Buyer - I & Purchasing Managers\\
    Dietary Aide / Server & Dining Room and Cafeteria Attendants and Bartender Helpers \\
    Customer Service Representative, San Francisco Airport & Customer Service Representatives \\
    Senior Software Engineer, Full Stack - Card Tech & Software Developers\\
    Senior Product Manager, Mulesoft Agent Fabric & Sales Engineers \\
    \bottomrule
    \end{tabular}
    \caption{Examples of titles from job postings mapped to the most similar O*NET occupation using our pipeline.}
    \label{tab:title_examples}
\end{table}

%% file: Figures/JobDescriptionExamples.tex
\begin{table*}[t]
\centering
\small
\begin{tabular}{p{0.30\textwidth} p{0.30\textwidth} p{0.30\textwidth}}
\toprule

\multicolumn{3}{l}{\textbf{Original job posting: Senior Project Engineer --- DPR Construction}} \\[2pt]

\multicolumn{3}{p{0.94\textwidth}}{
\textit{``DPR Construction is seeking a senior project engineer with a minimum
of 5 years of commercial construction experience... Project engineers will
work closely with all members of the project team and will be responsible
for the following: participation and application of DPR's environmental
health and safety plan... mentor, develop and train team members...
management of other project engineer team members... preparation,
monitoring, and updating project schedules... organizing and electronically
archiving project files...''}
} \\[6pt]

\midrule
\textbf{Extracted task statement} &
\textbf{Decomposed task} &
\textbf{Matched O*NET task} \\
\midrule

``Mentor, develop and train team members for fast-paced growth.''
& Mentor team members
& Supervise or train project team members, as necessary. \\[4pt]

&
Train team members
& Supervise or train project team members, as necessary. \\[4pt]

\midrule

``Preparation, monitoring, and updating project schedules...''
& Develop project schedules
& Develop or approve project plans, schedules, or budgets. \\[4pt]

&
Monitor project schedules against progress and constraints
& Oversee project production efforts to assure projects are completed on
time and within budget. \\[4pt]

\midrule

``Organizing and electronically archiving project files for easy access
by the project team.''
& Organize project files for efficient retrieval
& Maintain technical project files. \\

\bottomrule
\end{tabular}

\caption{We provide an example of our job-posting-to-O*NET mapping pipeline.
Starting from the full text of a job posting, we identify statements describing
work responsibilities, decompose each statement into atomic tasks, and map
each task to its corresponding O*NET task.}
\label{tab:task-mapping-example}
\end{table*}

%% file: Figures/Table/IWAShare.tex
\begin{table*}[t]
\centering
\footnotesize
\begin{tabularx}{\linewidth}{X r rr rr rr}
\toprule
 & & \multicolumn{2}{c}{Presence (\%)} & \multicolumn{2}{c}{AI-Linked Share (\%)} & \multicolumn{2}{c}{$\Delta$ Presence (pp)} \\
\cmidrule(lr){3-4} \cmidrule(lr){5-6} \cmidrule(lr){7-8}
IWA & AI posts. & 2023 & 2025 & 2023 & 2025 & AI & Other \\
\midrule
Develop scientific or mathematical theories or models & 144 & 1.9 & 1.8 & 31.5 & 38.0 & +0.10 & $-$0.15 \\
Design computer or information systems or applications & 109 & 9.9 & 7.2 & 2.5 & 9.8 & +0.47 & $-$3.08 \\
Program computer systems or production equipment & 70 & 6.1 & 4.0 & 3.1 & 10.7 & +0.24 & $-$2.31 \\
Analyze scientific or applied data using mathematical principles & 68 & 2.3 & 2.1 & 12.4 & 14.9 & +0.03 & $-$0.24 \\
Research technology designs or applications & 67 & 2.6 & 2.9 & 3.6 & 16.6 & +0.39 & $-$0.08 \\
Maintain current knowledge in area of expertise & 61 & 12.5 & 11.8 & 1.2 & 3.3 & +0.24 & $-$0.86 \\
Evaluate the characteristics, usefulness, or performance of products or technologies & 56 & 10.7 & 9.3 & 0.5 & 4.6 & +0.37 & $-$1.73 \\
Develop models of systems, processes, or products & 50 & 2.3 & 2.2 & 4.1 & 15.3 & +0.25 & $-$0.28 \\
Process digital or online data & 43 & 8.3 & 7.3 & 1.9 & 3.1 & +0.06 & $-$1.09 \\
Direct scientific or technical activities & 43 & 9.6 & 8.7 & 1.3 & 3.0 & +0.14 & $-$1.09 \\
Communicate with others about specifications or project details & 42 & 19.4 & 14.6 & 0.1 & 2.4 & +0.32 & $-$5.20 \\
Analyze business or financial data$^\dagger$ & 39 & 13.5 & 10.8 & 1.6 & 1.3 & $-$0.08 & $-$2.57 \\
Implement procedures or processes & 39 & 16.4 & 15.6 & 0.5 & 1.7 & +0.19 & $-$1.02 \\
Advise others on the design or use of technologies & 35 & 10.3 & 7.9 & 0.7 & 2.9 & +0.16 & $-$2.53 \\
Advise others on business or operational matters$^\dagger$ & 33 & 20.3 & 17.9 & 0.4 & 1.1 & +0.12 & $-$2.49 \\
\bottomrule
\end{tabularx}
\caption{The 15 IWAs that most frequently carry AI-related content. For each IWA, we report the total AI posts (number of postings in which the IWA is AI-linked) as well as the presence and AI-Linked Share in both 2023 and 2025. $\Delta$ AI and $\Delta$ Other report the change in the percentage of all postings containing the IWA with and without an AI link. $\dagger$ marks IWAs that the link audit identifies as largely mapping artifacts.}
\label{tab:ai_share_declining}
\end{table*}

%% file: Figures/FloorComp.tex
\begin{table*}[t]
\centering
\small
\begin{tabular}{llrrrrrr}
\toprule
DV & Threshold & Firms & Postings & Copilot $\beta$ & $p$ & Claude $\beta$ & $p$ \\
\midrule
\multirow{5}{*}{$\Delta$ Frequency} & $\geq 1$ & 30 & 25,590 & $-0.296$ (0.063) & $<$.001 & $-0.376$ (0.076) & $<$.001 \\
 & $\geq$50 & 28 & 25,504 & $-0.302$ (0.067) & $<$.001 & $-0.340$ (0.077) & $<$.001 \\
 & $\geq$100 & 23 & 24,261 & $-0.235$ (0.054) & $<$.001 & $-0.246$ (0.063) & $<$.001 \\
 & $\geq$200 & 20 & 22,963 & $-0.188$ (0.055) & $<$.001 & $-0.240$ (0.068) & $<$.001 \\
 & $\geq$250 & 17 & 21,285 & $-0.179$ (0.053) & $<$.001 & $-0.276$ (0.071) & $<$.001 \\
\midrule
\multirow{5}{*}{Log ratio} & $\geq 1$ & 30 & 25,590 & $-0.035$ (0.019) & .067 & $-0.043$ (0.017) & .010 \\
 & $\geq$50 & 28 & 25,504 & $-0.050$ (0.017) & .004 & $-0.049$ (0.017) & .004 \\
 & $\geq$100 & 23 & 24,261 & $-0.048$ (0.020) & .014 & $-0.039$ (0.018) & .028 \\
 & $\geq$200 & 20 & 22,963 & $-0.040$ (0.018) & .025 & $-0.036$ (0.019) & .056 \\
 & $\geq$250 & 17 & 21,285 & $-0.045$ (0.016) & .005 & $-0.046$ (0.020) & .024 \\
\midrule
\multirow{5}{*}{Mention share} & $\geq 1$ & 30 & 25,590 & $-0.021$ (0.007) & .004 & $-0.037$ (0.010) & $<$.001 \\
 & $\geq$50 & 28 & 25,504 & $-0.021$ (0.008) & .007 & $-0.033$ (0.011) & .002 \\
 & $\geq$100 & 23 & 24,261 & $-0.014$ (0.007) & .051 & $-0.022$ (0.010) & .028 \\
 & $\geq$200 & 20 & 22,963 & $-0.010$ (0.007) & .158 & $-0.024$ (0.011) & .026 \\
 & $\geq$250 & 17 & 21,285 & $-0.010$ (0.008) & .197 & $-0.031$ (0.011) & .007 \\
\bottomrule
\end{tabular}
\caption{Association between AI exposure and change in prevalence under different posting thresholds. The threshold is the minimum number of postings a firm must have in each window. The main analysis in Sec.~\ref{sec:prev} uses $\geq$200. The coefficient ($\beta$) is the fixed effect on standardized log AI exposure from a linear mixed-effects model with a random intercept and slope by firm; standard errors in parentheses. }
\label{tab:floor_robustness}
\end{table*}